# Modeling of cellular response after FLASH irradiation: a quantitative analysis based on the radiolytic oxygen depletion hypothesis

Hongyu Zhu[1,2,3], Junli Li [2,3*], Xiaowu Deng[1,*], Rui Qiu[2,3], Zhen Wu[2,4], Hui Zhang[2,3]

[1] Department of Radiation Oncology, Sun Yat-sen University Cancer Center, State Key Laboratory of Oncology in South China, Collaborative Innovation Center for Cancer Medicine, Guangzhou 510060, China [2] Department of Engineering Physics, Tsinghua University, Beijing 100084, China [3] Key Laboratory of Particle & Radiation Imaging (Tsinghua University), Ministry of Education, Beijing, China [4] Nuctech Company Limited, Beijing 100084, China

[*]Corresponding Author: dengxw@mail.sysu.edu.cn, lijunli@tsinghua.edu.cn

**Abstract**

**Purpose:** Recent studies suggest ultra-high dose rate (FLASH) irradiation can spare normal tissues from radiotoxicity, while efficiently controlling the tumor, and this is known as the "FLASH effect". This study performed theoretical analyses about the impact of radiolytic oxygen depletion (ROD) on the cellular responses after FLASH irradiation.

**Methods:** Monte Carlo simulation was used to model the ROD process, determine the DNA damage, and calculate the amount of oxygen depleted ($L_{ROD}$) during FLASH exposure. A mathematical model was applied to analyze oxygen tension ($pO_2$) distribution in human tissues and the recovery of $pO_2$ after FLASH irradiation. DNA damage and cell survival fractions (SFs) after FLASH irradiation were calculated. The impact of initial cellular $pO_2$, FLASH pulse number, pulse interval, and radiation quality of the source particles on ROD and subsequent cellular responses were systematically evaluated.

**Results:** The simulated electron $L_{ROD}$ range was 0.38–0.43 μM/Gy when $pO_2$ ranged from 7.5–160 mmHg. The calculated DNA damage and SFs show that radioprotective effect is only evident in cells with a lower $pO_2$. Different irradiation setups alter the cellular responses by modifying the $pO_2$. Single pulse delivery or multi-pulse delivery with pulse intervals shorter than 10–50 ms resulted in fewer DNA damages and higher SFs. Source particles with a low radiation quality have a higher capacity to deplete oxygen, and thus, lead to a more conspicuous radioprotective effect.

**Conclusions:** A systematic analysis of the cellular response following FLASH irradiation was performed to provided suggestions for future FLASH applications. The FLASH radioprotective effect due to ROD may only be observed in cells with a low $pO_2$. Single pulse delivery or multi-pulse delivery with short pulse intervals are suggested for FLASH irradiation to avoid oxygen tension recovery during pulse intervals. Source particles with low radiation quality are preferred for their conspicuous radioprotective effects.

# 1. Introduction

Ultra-high dose rate (FLASH) irradiation has attracted attention in the field of medical physics after promising results of several *in vivo* experiments (Vozenin *et al.*, 2019a; Montay-Gruel *et al.*, 2019; Favaudon *et al.*, 2014) reported in recent years, which have shown that FLASH can spare healthy tissues, while efficiently controlling tumor, a phenomena referred to as the FLASH effect (Wilson *et al.*, 2020; Vozenin *et al.*, 2019b). FLASH irradiation refers generally to have a mean dose rate of > 40 Gy/s and instantaneous dose rate of > $10^5$ Gy/s, and this is of 3-4 magnitude higher than conventional dose rate (CONV) irradiation adopted in clinical applications. Although the underlying mechanism of the FLASH effect remains unclear, the successful treatment of the first cancer patient using FLASH radiotherapy (Bourhis *et al.*, 2019) makes it a promising technique for clinical application.

Several non-mutually exclusive hypotheses have been proposed to explain the FLASH effect, including those based on the modified transforming growth factor beta (TGF-β) signaling (Buonanno *et al.*, 2019; Favaudon *et al.*, 2014), a reduced dose of immune cells in the circulating blood system (Durante *et al.*, 2018; Jin *et al.*, 2020), and radiolytic oxygen depletion (ROD) (Montay-Gruel *et al.*, 2019; Spitz *et al.*, 2019; Pratx and Kapp, 2019a) during FLASH irradiation. The ROD hypothesis, the leading hypothesis to explain FLASH effect, was first proposed to explain the *in vitro* experimental results obtained in the earlier studies(Town, 1967; Berry *et al.*, 1969). When irradiated with a single pulse beam that lasted for nanoseconds (ns) to microseconds (μs), cells were found to be more radioresistant than when exposed to CONV irradiation (e.g., one administered using $^{60}$Co) (Berry *et al.*, 1969). These results were explained by the rapid depletion of oxygen during FLASH irradiation, which is not possible under CONV irradiation, and by creating a local radiobiological hypoxic environment, thus making cells less radiosensitive. Some other studies reported that FLASH pulses induce a reduced radiosensitivity only in cells with a low level of oxygen (Berry and Stedeford, 1972; Adrian *et al.*, 2020). More recently, Montay-Gruel *et al*. found the neurocognitive benefits of FLASH irradiation in mice whole-brain irradiation experiments, and the benefits were reversed when the oxygen concentration was doubled in the brain during FLASH irradiation. These results were explained by the hypothesis that the yield of reactive oxygen species was reduced due to the rapid ROD in the brain during FLASH irradiation (Montay-Gruel *et al.*, 2019).

Oxygen is regarded as one of the key factors that modifies radiobiological responses, and experimental results have shown that hypoxic cells are more radioresistant than their aerobic counterparts (Hall and Giaccia, 2018). Oxygen-dependent radiosensitivity is usually explained by the oxygen fixation hypothesis (OFH), which considers that ionizing radiation can produce DNA radicals via chemical reactions, DNA radicals can be chemically restored by thiols in the cellular environment or react with dissolved oxygen molecules and form DNA peroxyl radicals, which are difficult to be repaired and can be regarded as fixed damage. Therefore, hypoxic cells suffer less DNA damage after irradiation (Hall and Giaccia, 2018; Liu *et al.*, 2015).

Monte Carlo track structure (MCTS) codes can simulate the detailed physics tracks and chemical reactions of primary particles at the nanometer (nm) scale, and have been extensively applied in cellular and sub-cellular radiobiological response modeling under CONV irradiation (Meylan *et al.*, 2017; Friedland *et al.*, 2017; Zhu *et al.*, 2020). Current MCTS codes simulate particle tracks independently, which is realistic under low dose rate irradiation; however, incident particles are densely distributed in spatial and temporal proximity under FLASH irradiation, and intertrack reactions should be considered in FLASH simulations and such simulation can be very time consuming. Some recent progresses were made to adapt MCTS codes for FLASH simulation. Ramos-Méndez *et al*. incorporated the independent reaction time (IRT) method in the TOPAS-nBio code to study the intertrack effects in the chemical stage of water radiolysis under FLASH irradiation (Ramos-Méndez *et al.*, 2020). Lai *et al*. implemented oxygen related chemical reactions in the graphical processing unit (GPU)-based gMicroMC code and applied it for the study of ROD hypothesis of FLASH irradiation, the oxygen enhancement ratio (OER) was calculated to quantify changes in biological effects due to changes in oxygen concentration (Lai *et al.*, 2020). Further study of the FLASH induced chemical reactions and cellular responses should be performed.

In this study, the impact of ROD on cellular responses after FLASH irradiation was quantitatively analyzed using Monte Carlo (MC) simulations and mathematical modeling. The MCTS code NASIC (Nanodosimetry Monte Carlo Simulation Code) was adapted to the FLASH simulations, and the oxygen depletion process and information related to DNA damage under different irradiation setups were obtained. The recovery of oxygen tension ($pO_2$, in the unit of mmHg) in cells due to oxygen diffusion from vessels was described using differential equations and solved using a numerical method. The impact of initial oxygen tension, FLASH pulse structure, and radiation quality on the cellular responses were systematically analyzed in this work.

## 2. Methods and materials

### 2.1 Monte Carlo simulation of FLASH pulse

Currently, almost all published MCTS simulation studies of radiobiological responses have been performed using the CONV irradiation setup and adopted the assumption that particle tracks are independent of each other (Meylan *et al.*, 2017; Friedland *et al.*, 2017; Zhu *et al.*, 2020). Simulations of the physical stage (< $10^{-15}$ s), the pre-chemical stage ($10^{-15}$–$10^{-12}$ s), and the chemical stage ($10^{-12}$–$10^{-6}$ s) were performed sequentially for each primary particle. For the modeling of DNA damage, the chemical stage was terminated within a few nanoseconds (Zhu *et al.*, 2020; Meylan *et al.*, 2017; Friedland *et al.*, 2017; Friedland *et al.*, 2003; Nikjoo *et al.*, 2001) because it is assumed that the indirect DNA damage results from interactions between the DNA structure and the hydroxyl radicals (˙OH), which have a lifetime of a few nanoseconds in the cellular environment (Roots and Okada, 1975).

Such a simulation method cannot be applied to FLASH simulation owing to two main differences between FLASH irradiation and CONV irradiation. First, a tremendously high number of particles are emitted in a single short pulse (ns to μs scale) and the particle tracks are highly overlapped. Though the intertrack physical reactions can be neglected (Kreipl *et al.*, 2009; Ramos-Méndez *et al.*, 2020) because the temporal separation between the particle tracks is much greater than the duration of the physical stage (Ramos-Méndez *et al.*, 2020), the track overlap can have a significant impact on the chemical reactions because chemical radicals generated by different primary particles can react among themselves. Kreipl *et al.* investigated the impact of spatial and temporal track overlap on the chemical radical yields, and their results showed that the radical yields were considerably modified by intertrack reactions (Kreipl *et al.*, 2009).

Second, $e_{aq}^-$ and $\cdot$H radicals, which are the main $O_2$ scavengers, are produced in response to the FLASH pulse at very high instantaneous concentrations; $O_2$ is rapidly depleted during FLASH irradiation due to the chemical reactions of Eq. 1–2 (Nikjoo *et al.*, 2006; Boscolo *et al.*, 2020; Spitz *et al.*, 2019), leading to a significant change in the oxygen tension in the biological tissues, followed by modified radiobiological response. Therefore, the change in oxygen concentration should be carefully recorded during FLASH irradiation to model the biological response, and the chemical stage should be tracked for the entire lifetime of $e_{aq}^-$ and $\cdot$H. Roots *et al.* measured the lifetime of $\cdot$OH, $e_{aq}^-$, and $\cdot$H in the cellular environment, which was reported to be 1.6–4.4 ns, 0.19–4 μs, and 40–170 μs, respectively (Roots and Okada, 1975).

$$\cdot H + O_2 \rightarrow HO_2 \tag{1}$$

$$e_{aq}^- + O_2 \rightarrow O_2^- \tag{2}$$

In this study, the MCTS code NASIC (Chen *et al.*, 2017; Li *et al.*, 2015) was used for MC simulations of FLASH irradiation. NASIC has already been applied to simulations of radiobiological responses, including DNA damage and the radiosensitization effect of gold nanoparticles under CONV irradiation (Chen *et al.*, 2017; Li *et al.*, 2015; Xie *et al.*, 2013; Li *et al.*, 2020); NASIC was further modified for simulations under FLASH irradiation.

In the physics stage, the track structure of *N* primary particles that can transverse the cell nucleus during a FLASH pulse were simulated independently, and all the energy depositions induced by primary particles and their secondary particles were recorded as a track structure file. This file was used as the input file for the simulation of the pre-chemical stage to generate sub-excited electrons, excited water molecules, and ionized water molecules, and simulate the following radiolysis process. In this manner, the initial chemical radicals produced by different primary particles in the pulse could be considered in the same pool, and thus, their interactions with each other could be simulated.

The chemistry module of NASIC was modified to support the simulation in oxygenated water. The lists of chemical radicals and chemical reactions were extended according to Nikjoo *et al.* (Nikjoo *et al.*, 2006). As directly introducing oxygen molecules in MC simulations increases the time costs dramatically, an approximation method (Boscolo *et al.*, 2020; Colliaux *et al.*, 2015; Tomita *et al.*, 1997; Pimblott *et al.*, 1991; Green, 1984) was adopted in this work to simulate oxygen depletion reactions. The dissolved oxygen is regarded a homogeneous continuum, and the probability that a chemical radical reacts with the oxygenated continuum in time $t$ can be calculated as

$$P(t) = 1 - \exp\left(-4\pi D_{sum} R_C C_s \left[t + 2R_C \sqrt{\frac{t}{\pi D_{sum}}}\right]\right) \qquad (3)$$

where $D_{sum}$ is the sum of the diffusion coefficients of oxygen and the radical, $R_C$ is the reaction radius between oxygen and the radical, $C_s$ is the concentration of dissolved oxygen (mol/L, M), and it was calculated with $C_s = $ pO$_2 \times H_C$, here, $H_C$ = 1.73 μM/mmHg is the coefficient of Henry's Law for oxygen dissolving in water (Boscolo *et al.*, 2020; Colliaux *et al.*, 2015). If the oxygen depletion reaction occurs, then the radical is replaced by the corresponding reaction product; for example, in a reaction between $e_{aq}^-$ and O$_2$, $e_{aq}^-$ is replaced by O$_2^-$ (please see the supplementary material section S1 and S2 for more details).

The impact of radical life time on the amount of oxygen depleted in FLASH irradiation ($L_{ROD}$, in units of μM/Gy) was carefully studied, and the detailed analysis can be found in the supplementary material section S3. In this work, the lifetime values of ·H, $e_{aq}^-$, and ·OH were set to 1 μs, 40 μs, and 1 ns, respectively, for simulations of FLASH irradiation.

## 2.2 Mathematic modeling of oxygen diffusion after FLASH irradiation

### 2.2.1 Oxygen tension distribution in tissues at steady conditions

The mathematical model developed by Grimes *et al.* (Grimes *et al.*, 2014) was adopted in this work to describe the oxygen tension distribution in tissues at a steady state; that is, the state in which the tissues are not irradiated, or long enough after FLASH irradiation, and the oxygen tension is already recovered. The blood vessel was simplified as having a cylindrical geometry with a radius of $r_0$, and oxygen diffusion from the blood vessels was considered to occur in a direction perpendicular to the vessel walls, being able to reach a maximum distance of $r_c$; thus, the pO₂ distribution for $r_0 \leq r \leq r_c$ at a steady state, $p(r)$, can be described as:

$$0 = D_{O_2} \frac{1}{r} \frac{1}{dr}\left(r \frac{dp}{dr}\right) - C_{meta} \tag{4}$$

wherein, $D_{O_2} = 2.1 \times 10^{-9}$ m²/s is the diffusion rate of oxygen (Nikjoo *et al.*, 2006), and $C_{meta} = 15$ mmHg/s is the consumption rate of oxygen due to metabolism (Grimes *et al.*, 2014), and Eq. 4 has two boundary conditions of $p(r = r_0) = p_0$ and $p'(r_c) = 0$; thus, the analytic solution of Eq. 4 is given by:

$$p(r) = p_0 + \frac{C_{meta}}{4D_{O_2}}\left(r^2 - r_0^2 - 2r_c^2 \ln\frac{r}{r_0}\right) \tag{5}$$

### 2.2.2 Oxygen tension recovery after FLASH irradiation

The oxygen tension in tissues can be changed due to ROD of FLASH irradiation and will gradually recover after irradiation due to oxygen diffusion from vessels, and the dynamic change of pO₂ in the recovery process can be described as:

$$\frac{dp}{dt} = D_{O_2} \frac{1}{r} \frac{1}{dr}\left(r \frac{dp}{dr}\right) - C_{meta} \tag{6}$$

Eq. 6 has two boundary conditions: the initial pO₂ distribution immediately after the FLASH pulse and the final pO₂ distribution after a full recovery ; the initial pO₂ distribution is the pO₂ distribution described by Eq. 5 minus the amount of oxygen depleted during irradiation; and the final pO$_2$ distribution is the pO$_2$ distribution at steady state described by Eq. 5. MATLAB (Mathworks Inc.) was used to solve Eq. 6 using the finite differential method.

**2.3 Modeling of cellular response after irradiation**

In this work, two biological endpoints, the DNA damage yields and cell survival fractions (SFs) of the Chinese hamster ovary (CHO) cell line at different initial oxygen levels after CONV and FLASH irradiation were calculated and compared. The FLASH pulse was assumed to have an ultra-short pulse length and primary particles were assumed to incident instantaneously, therefore the intertrack chemical reactions were considered for FLASH pulse. The primary particles of CONV irradiation were considered to be separated, and they were simulated independently. DNA damage yield was calculated with the previously developed and well-validated DICOLDD (different cell oxygen level DNA damage) model (Zhu *et al.*, 2021); the DICOLDD model was developed based on OFH, and can be applied for DNA damage calculation under different cellular oxygen tension; Cell SFs were calculated using a mechanistic cellular survival model developed by Wang *et al*. (Wang *et al.*, 2018), which connected the relationship between radiation-induced double strand break (DSB) in the nucleus and the probability of cell survival. A brief introduction of the DICOLDD model and the mechanistic cellular survival model can be found in the supplementary material section S4 and S5, respectively.

An index of radiosensitivity ratio (RSR) was defined to quantify the impact of ROD in FLASH. RSR was calculated as the ratio between doses under FLASH irradiation and under CONV irradiation needed to achieve the same biological effect (e.g., a given level of DNA damage or cell killing), and RSR describes the increment of dose tolerance due to oxygen depletion under FLASH irradiation.

$$\text{RSR} = \frac{D_{\text{FLASH}}}{D_{\text{CONV}}}|_{\text{isoeffect}} \tag{7}$$

## 2.4 Impact of initial oxygen tension and FLASH pulse delivery strategy on cellular responses

The impact of initial cell oxygen tension, FLASH pulse number, FLASH pulse interval, and radiation quality of source particles on the cellular response (DNA damage yields and cell survival fractions) was investigated and compared.

Our previous calculation (Zhu *et al.*, 2021) showed that the change in oxygen concentration (OC) only had a limited impact on the DNA damage yield when OC was higher than 5%. However, when OC decreased below 1%, the DNA damage yield changed significantly with the OC. Experimental results (Berry and Stedeford, 1972; Adrian *et al.*, 2020) and computational predictions (Petersson *et al.*, 2020; Pratx and Kapp, 2019b; Pratx and Kapp, 2019a) showed that the FLASH effect may only be observed in cells at a relatively low oxygen tension (< 15 mmHg). Adrian *et al.* measured SFs of prostate cancer cells at different oxygen levels (OC = 1.6% - 20%) after CONV and FLASH irradiation. Their experimental results showed no difference between FLASH and CONV under normoxic conditions; however, a significant increase in SF after FLASH irradiation was found in the high dose region (18 Gy) in hypoxic cells (OC = 1.6%) (Adrian *et al.*, 2020). In this work, DNA damage yields under CONV and FLASH irradiation were calculated with cells at an initial $pO_2$ of 3–15 mmHg (OC = 0.4%–2%), and SFs of CHO cells after CONV and FLASH irradiation were calculated using cells at an initial $pO_2$ of 7.5–160 mmHg (OC = 1%–21%). The considered source particle was a 4.5 MeV electron, which is a usually used beam source in many FLASH experiments (Vozenin *et al.*, 2019a; Montay-Gruel *et al.*, 2017; Favaudon *et al.*, 2014).

To investigate the impact of FLASH pulse number on the cellular response, 10 Gy of dose were delivered in 1, 2, 5, or 10 pulses in 1 second. The impact of the FLASH pulse interval was also studied by delivering a total dose of 10 Gy in 10 pulses, and the pulse interval was set to 10, 25, 50 200, and 1,000 ms. In addition, the impact of radiation quality was examined by setting the beam source as a 4.5 MeV electron (LET = 0.3 keV/μm), a 100 MeV proton (LET = 0.59 keV/μm), a 4.5 MeV proton (LET = 9.3 keV/μm), and a 1 MeV proton (LET = 32.4 keV/μm).

## 3. Results

### 3.1 Oxygen depletion during FLASH irradiation

$L_{ROD}$ was simulated in cells with a different initial pO$_2$, and a typical FLASH pulse (4.5 MeV electron with a total dose of 10 Gy) was considered. Simulated $L_{ROD}$ increased with the initial pO$_2$ of the tissue, with $L_{ROD}$ = 0.38, 0.39, 0.41, and 0.43 µM/Gy (equivalent to 0.22, 0.23, 0.24, and 0.25 mmHg/Gy), when pO$_2$ = 7.5, 15, 37.5 and 160 mmHg, respectively .

$L_{ROD}$ can be experimentally measured using solutions pre-equilibrated at a certain level of oxygen, and then sealed with glass to avoid gas exchanges during irradiation. Weiss *et al*. measured oxygen depletion using solutions (equilibrated at 30 mmHg and 160 mmHg O$_2$) exposed to a beam of 0.45 MeV electrons in single pulses of about 3 ns and reported an $L_{ROD}$ = 0.58±0.1 µM/Gy (Weiss *et al.*, 1974). Michaels *et al*. measured oxygen depletion using solutions irradiated with 25 MV X-rays and reported an $L_{ROD}$ = 0.44 µM/Gy (Michaels, 1986). The $L_{ROD}$ values simulated here are in line with the experimental results.

### 3.2 Oxygen distribution in biological tissues

Eq. 5 was used to calculate the pO$_2$ distribution in the tissues around the vessels at a steady state, and the pO$_2$ within the vessel was set to $p_0$ = 40 mmHg for the veins and $p_0$ = 100 mmHg for the arteries. Following previously reported definitions (Hall and Giaccia, 2018; Liu *et al.*, 2015), we delimited radiobiological hypoxia as pO$_2$ < 3 mmHg and pathological hypoxia as 3 mmHg < pO$_2$ < 15 mmHg. **Figure 1** shows that tissues that are 48 µm away from the center of the artery are exposed to pathological hypoxia, those that are 77 µm away from the center of the artery experience radiobiological hypoxia, those that are 20 µm away from the center of the vein experience pathological hypoxia, and those 45 µm away from the center of the vein experience radiobiological hypoxia.

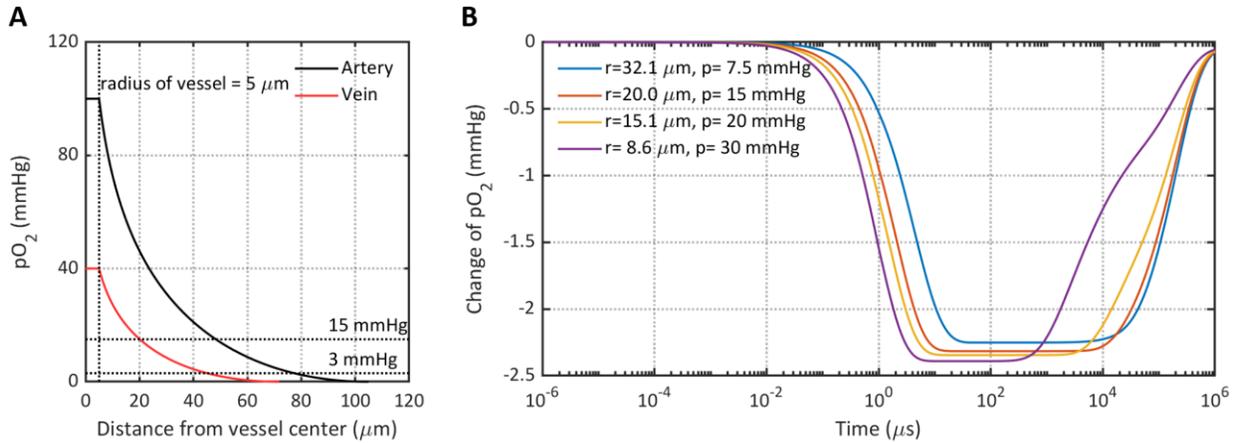

*Figure 1.(A) pO$_2$ distribution in tissues around the artery and vein vessels. The vessel radius was set to 5 μm and a p$_0$ =100 mmHg and a p$_0$ =40 mmHg were assumed for the artery and vein vessels, respectively. (B) Dynamic change in pO$_2$ in the tissues around the vein after being exposed to 4.5 MeV single FLASH pulse with a total dose of 10 Gy. Curves in different colors represent the tissues located at different distances from the vein center (pO$_2$ = 40 mmHg within vessel), curve labels represent the distance, r, to the vein vessel center and the initial pO$_2$, p, at a steady state.*

**Figure 1B** shows the dynamic change of pO$_2$ in the tissues around the vein (pO$_2$ = 40 mmHg within the vessel) after irradiation by a 4.5 MeV electron single FLASH pulse with a total dose of 10 Gy. As seen, tissues closer to the vein center show a more rapid decrease in pO$_2$ after irradiation due to a higher initial pO$_2$, and a faster recovery of pO$_2$ due to a faster oxygen diffusion supplement from the vessels. The pO$_2$ values dropped to the lowest level within 20 μs due to oxygen depletion. The low pO$_2$ level was maintained for 1–10 ms, and the oxygen tension in tissues was generally fully recovered within 1 second.

### 3.3 Impact of initial oxygen level on the cellular response

In the following sections, the cellular responses to 4.5 MeV electrons delivered using CONV of FLASH mode were considered. The DNA damage yield induced by 4.5 MeV electrons was simulated using NASIC and the initial DSB radical yield after radiation was $n_0 = 5.0$ Gy$^{-1}$Gbp$^{-1}$, and the initial direct DSB yield after irradiation was $N_{dir} = 2.8$ Gy$^{-1}$Gbp$^{-1}$. Owing to the ROD in FLASH irradiation, the oxygen tension in cells decreased after FLASH irradiation, leading to a reduced indirect DSB yield (Eq. S7–Eq. S10 in the supplementary material) compared with CONV irradiation.

A significant difference was observed in indirect damage yield induced by CONV and FLASH irradiation in cells with a lower initial oxygen tension. For cells with an initial oxygen tension of $pO_2 = 15$ mmHg, the indirect DSB yields induced by FLASH irradiation and CONV irradiation were 3.40 and 3.57 $Gy^{-1}Gbp^{-1}$, respectively, with a relative difference of 4.8% (= 1-3.40/3.57); whereas for cells with an initial oxygen tension of $pO_2 = 3$ mmHg, the indirect DSB yield induced by FLASH irradiation and CONV irradiation were 0.86 and 1.71 $Gy^{-1}Gbp^{-1}$, respectively, with a relative difference of approximately 49.7%.

Reduced DNA damage after FLASH irradiation can result in higher cell SFs, and **Figure 2** shows the calculated SFs (Eq. S11–Eq. S12 in the supplementary material) of CHO cells at a different initial oxygen level that were exposed to 4.5 MeV electrons delivered using CONV (dashed lines) or FLASH (solid lines) mode. Although a different cell line was considered in our calculation, our results are in line with those of Adrian *et al*. (Adrian *et al.*, 2020). For CHO cells at a higher oxygen tension (≥ 37.5 mmHg), no difference in SFs was observed between CONV and FLASH. In contrast, increased SFs are observed under FLASH irradiation in hypoxic cells, and an enlarged difference in SFs between CONV and FLASH irradiation was observed with increasing dose.

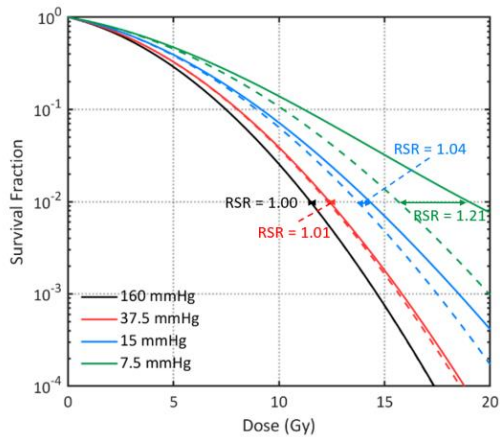

*Figure 2. Survival fractions of CHO cells with different initial $pO_2$ values after being exposed to 4.5 MeV electron CONV irradiation (dashed line) or single FLASH pulse (solid line).*

To quantify difference in cell SFs after CONV and FLASH irradiation, RSR values were calculated using Eq. 7 to describe the increased dose tolerance after FLASH irradiation to reach the endpoint of SF = 1%.

RSR values were labeled in **Figure 2**. For CHO cells at a relatively high oxygen tension (> 37.5 mmHg), RSR were very close to 1, and no increase in dose tolerance after FLASH irradiation was observed. For CHO cells with an initial pO$_2$ of 7.5 mmHg, RSR increased to 1.21, which means that a 21% higher dose is required under FLASH irradiation to reach the endpoint of SF = 1%, compared with CONV irradiation.

## 3.4 Impact of FLASH pulse characteristic (pulse number and pulse interval) on the cellular response

To investigate the impact of FLASH pulse fractionation on the cellular response, CHO cells (initial pO$_2$ = 7.5 mmHg) were exposed to 4.5 MeV electron FLASH irradiation with a total dose of 10 Gy delivered in 1, 2, 5, and 10 pulses. **Figure 3A** shows the dynamic changes in cellular pO$_2$ within 1 second; pO$_2$ values decrease rapidly during FLASH irradiation and gradually recover after irradiation. When cells were irradiated in the multi-pulse irradiation mode, the cell oxygen tension recovered partially during the pulse interval, and the lowest pO$_2$ level after a single FLASH pulse radiation could not be reached, which led to differential cellular responses. **Table 1** summarizes the responses of CHO cells after irradiation; the average value of pO$_2$ after a single pulse was 21% lower than that after 5 pulses, and this further led to a 5.4% lower indirect DNA damage yield and a 16% higher SF.

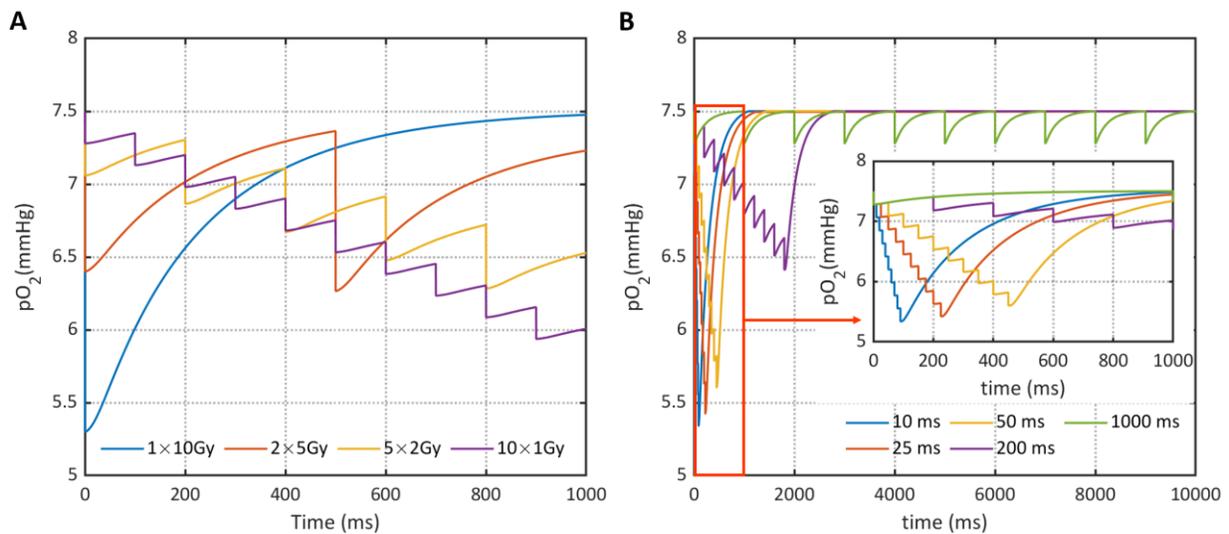

*Figure 3. Change of pO$_2$ in cells with an initial pO$_2$ of 7.5 mmHg exposed to 4.5 MeV electron FLASH irradiation. (A)*

*A total dose of 10 Gy delivered in 1, 2, 5, or 10 pulses in 1 second. (B) 10 Gy delivered in 10 pulses which were separated by different intervals.*

*Table 1. The average $pO_2$, total DNA damage yield, and cell survival fractions of CHO cells (initial $pO_2$ = 7.5 mmHg) were exposed to 4.5 MeV electron FLASH irradiation with a total dose of 10 Gy delivered in 1, 2, 5, and 10 pulses; or 10 Gy delivered in 10 pulses which were separated by different intervals.*

| Number of pulses delivering 10 Gy | Average $pO_2$ after irradiation (mmHg) | Total DSB yield ($Gbp^{-1} 10 Gy^{-1}$) | SFs |
|---|---|---|---|
| 1 | 5.3 | 51.94 | 0.140 |
| 2 | 6.33 | 54.13 | 0.122 |
| 5 | 6.67 | 54.75 | 0.118 |
| 10 | 6.61 | 54.61 | 0.119 |
| Maximum relative difference * | 21% | 5.1% | 15.7% |
| Pulse interval (ms) | Average $pO_2$ after irradiation (mmHg) | Total DSB yield ($Gbp^{-1} 10 Gy^{-1}$) | SFs |
| 10 | 6.31 | 54.02 | 0.123 |
| 25 | 6.35 | 54.10 | 0.122 |
| 50 | 6.44 | 54.28 | 0.121 |
| 200 | 6.85 | 55.06 | 0.115 |
| 1,000 | 7.28 | 55.82 | 0.110 |
| Maximum relative difference * | 13% | 3.2% | 10.6% |

* The maximum relative difference was calculated using: (1 - minimum/maximum.) × 100%

As shown in **Figure 1B**, the low $pO_2$ level caused by ROD was maintained for milliseconds to tens of milliseconds, after which $pO_2$ started to recover. Different pulse intervals result in different initial $pO_2$ for the next pulse and changes the cellular responses accordingly. The impact of pulse interval on the cellular response was qualified by irradiating CHO cells (initial pO$_2$ = 7.5 mmHg) with 10 × 1 Gy of 4.5 MeV electron FLASH pulses, and each pulse was separated by 10, 25, 50, 200, and 1,000 ms; the dynamic change of is shown in **Figure 3B**. Pulse interval of lower than 50 ms had a limited impact (< 2%) on the cellular responses, but the cell SF decreased significantly when the pulse interval was extended to 1,000 ms, due to the full recovery of oxygen tension during pulse intervals.

A similar analysis for the impact of pulse fractionation and pulse interval on the cellular response was performed for CHO cells with an initial pO$_2$ of 15 mmHg, and the results have been provided in Table S5.

### 3.5 Impact of radiation quality on the cellular response

Radiation quality can modify the cellular responses after FLASH irradiation in different ways; on one hand, high LET radiations can lead to higher DNA damage yields and induce more complex DNA damage distributions that are less likely to be repaired accurately and may result in chromosomal aberrations and lead to cell death. On the other hand, high LET radiation deposits more energy locally and produces chemical radicals with a denser spatial distribution, which makes it easier for radicals to react with each other and reduces the probability of reactions between dissolved oxygen molecules and e$_{aq}^-$, as well as $\cdot$H; thus, a high LET radiation has lower oxygen depletion capability and the difference between the cellular responses induced by a high LET radiation in CONV and FLASH mode would also be reduced.

**Figure 4** shows $L_{ROD}$ values increase with pO$_2$, but decrease with radiation quality. The $L_{ROD}$ of 4.5 MeV electrons (LET = 0.3 keV/μm) reached 0.43 μM/Gy (0.25 mmHg/ Gy) at 760 mmHg, which was 38.9% higher than that of the 1.0 MeV protons (LET = 32.4 keV/μm). **Figure 4B** shows low LET radiations show enlarged increments of SFs after FLASH irradiation, compared with a high LET radiation. The RSR value was 1.03 for 1.0 MeV protons, whereas it increased to 1.21 for 4.5 MeV electrons when an SF = 1% was adopted as the endpoint, indicating that a low LET radiation shows a better FLASH sparing effect on cell survival.

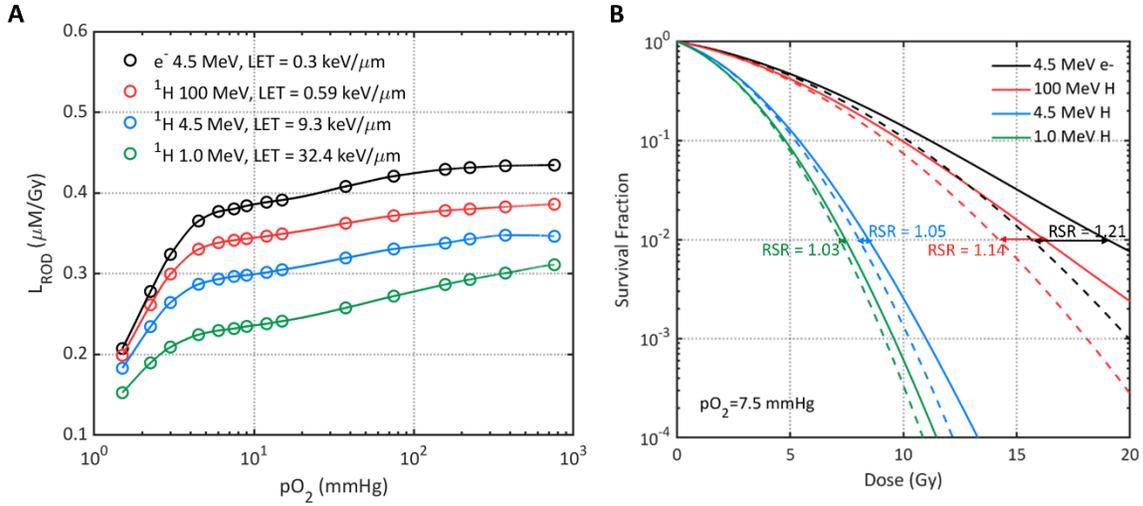

*Figure 4. (A) The relationship between $L_{ROD}$ and $pO_2$ for different source particles. (B) Survival fractions of CHO cells with an initial $pO_2$ of 7.5 mmHg after being exposed to CONV (dashed lines) single FLASH pulses (solid lines) with different radiation quality.*

## 4. Discussion

In this study, we used MC simulations and analytical calculations to investigate the modified cellular responses in hypoxic normal cells under FLASH irradiation due to ROD, and our results show that at lower the initial oxygen tension, a more evident modified cellular response (higher RSR value, which means a more obvious radioprotective effect) can be observed after FLASH irradiation. One may deduce that hypoxic cancer cells are also prone to having a radioprotective effect after FLASH irradiation, and this deduction seems to contradict the published experimental results, which have shown that FLASH irradiation is as effective as CONV irradiation in killing tumor targets.

It should be noted that, according to the simulated and experimental $L_{ROD}$, when human tissues ($pO_2$ = 1–100 mmHg, corresponding to $C_S$=1.73 – 173 μM) are exposed to a typical FLASH pulse (4.5 MeV electron with a total dose of 10 Gy), less than 4.29 μM oxygen can be depleted during irradiation. Previous studies have shown that oxygen tension has a very limited impact on the cellular response when $pO_2$ is higher than 30–40 mmHg (Hall and Giaccia, 2018); however, the DNA damage yield (Zhu *et al.*, 2021) and cell SF (Ling *et al.*, 1981) changes significantly when pO$_2$ is lower than 7.5 mmHg. This indicates that the change in cellular response due to oxygen depletion under FLASH irradiation can only be observed in cells at a relatively low oxygen level, such as cells located far away from the vessel center or hypoxic stem cell niches in normal tissues. Studies have shown that stem cell niches in the bone marrow experience local oxygen tension of 9.9–32 mmHg (Spencer *et al.*, 2014), that the pO$_2$ of mesenchymal stem cells ranges from 15 to 60 mmHg, and the pO$_2$ of hematopoietic stem cells ranges from 7 to 50 mmHg (Mohyeldin *et al.*, 2010). Pratx and Kapp proposed that hypoxic stem cells may be related to radioprotection from ROD associated with the FLASH effect (Pratx and Kapp, 2019b; Pratx and Kapp, 2019a). Therefore, in this work, we focused on the cellular response of hypoxic cells.

The mechanism underlying the FLASH effect remains unclear and many different factors, such as the immunological signaling, modified microenvironment, and ROD effect, may act collectively to affect the FLASH effect (Zhou, 2020). In this work, we only investigated the impact of ROD on the cellular responses after FLASH irradiation and other possible influencing factors were not considered. Therefore, the radioprotective effect resulting in ROD may be canceled out or enhanced by other factors in realistic biological tissues. For instance, the differential response to reactive oxygen species (ROS) between tumor cells and normal cells can result in different cellular responses. ROS, including $O_2^-$, $H_2O_2$, and ·OH, are byproducts of aerobic metabolism. Low to moderate levels of endogenous ROS are required for cell proliferation, differentiation, and survival, and a mild increase in the ROS levels may result in a transient alteration in these activities; however, when the ROS level exceeds the threshold level, it may overwhelm the antioxidant capacity and result in cell death (Trachootham *et al.*, 2009). Cancer cells have higher levels of endogenous ROS compared to their normal counterparts due to the increased metabolism. When the redox balance in cells is disturbed, for example, when cells are exposed to FLASH irradiation, enormous amounts of ROS are produced due to radiochemical reactions (Labarbe *et al.*, 2020; Spitz *et al.*, 2019; Montay-Gruel *et al.*, 2019). It is easier for cancer cells to reach the ROS threshold that can selectively kill them but spare normal cells. Therefore, cancer cells are protected by ROD but are selectively damaged due to excessive ROS generation after FLASH irradiation. These two factors may cancel out the radioprotective effects of ROD.

## 5. Conclusions

In this work, the impact of ROD on the cellular response under FLASH irradiation was studied using MC simulations and mathematical modeling. The MCTS code NASIC was updated to support the simulation of oxygen depletion reactions. The calculated $L_{ROD}$ values in a normal human tissue ($pO_2$ = 1–100 mmHg) were lower than 0.43 μM/Gy (0.25 mmHg/Gy), which indicated that the modified cellular response due to oxygen depletion in FLASH irradiation can only be observed in hypoxic cells ($pO_2$ < 30–40 mmHg).

We have systematically evaluated the impact of the initial oxygen tension, FLASH pulse number, pulse interval, and radiation quality of source particles on the cellular response. An index of RSR was proposed in this study to quantify the difference in cellular response after CONV and FLASH irradiation. Based on our calculations, we infer the following:

1. The FLASH radioprotective effect due to ROD is only evident in cells with a low $pO_2$.
2. The multi-pulse delivery strategy is not preferred for obtaining a better radioprotective effect under FLASH irradiation because the cellular $pO_2$ can be partially, or completely, recovered during pulse interval.
3. If the single pulse delivery regimen is not available in the practical application of FLASH irradiation, a pulse interval of less than 10–50 ms is suggested to take advantage of the oxygen tension preservation window and superpose the ROD caused by different pulses.
4. Low LET source particles, such as electrons, X-rays, and high-energy protons are preferred in FLASH irradiation to provide a better radioprotective effect to normal cells.

Our conclusions can serve as a reference for future FLASH applications, and the method developed in this work is applicable for prediction and comparison of future *in vitro* experimental results. Analyses that incorporate ROD, as well as other possible influencing factors, should be performed to better understand the FLASH effect. More experimental and theoretical analyses are needed to ascertain the biological mechanism of FLASH and promote the clinical application of FLASH radiotherapy.

**Acknowledgements**

This work was funded by the National Key R&D Program of China (2017YFC0113200), Cancer Precision Radiotherapy Spark Program of China International Medical Foundation (2019-N-11-20), and National Natural Science Foundation of China (12005316).

# Modeling of cellular response after FLASH irradiation: a quantitative analysis based on radiolytic oxygen depletion hypothesis

**Supplementary material**

### S1. Simulation algorithm for chemical reactions in NASIC

In the chemical stage, NASIC uses the step-by-step algorithm, which is a generally adopted method in Monte Carlo track structure (MCTS) codes[1-4], for the simulation of diffusion and reactions between chemical radicals. In each step, all chemical radicals were assumed to diffuse with a randomly selected direction and the displacement of radicals is sampled with:

$$\sqrt{\langle r^2 \rangle} = \sqrt{6D\Delta t} \qquad (S1)$$

Where, $\sqrt{\langle r^2 \rangle}$ is the root mean square (RMS) displacement, $D$ is the diffusion constant (see Table S1), $\Delta t$ is the time step. After the diffusion of radicals, the distance between each pair of radicals was scanned, if the spatial distance of two reaction partners is less than the corresponding reaction radius, a chemical reaction is assumed to happen, and the reaction radius, $R_C$, can be calculated with:

$$k_{obs} = 4\pi D_{sum} R_C \frac{R_C}{R_C + \sqrt{\pi D_{sum} \Delta t}} \qquad (S2)$$

Where, $k_{obs}$ is the observed reaction rate constant (see Table S2), $D_{sum}$ is the sum of diffusion constants of two reactants.

Table S1. List of chemical species and corresponding diffusion coefficients ($10^{-9}$ m$^2$s$^{-1}$) adopted in different MCTS codes

| ID | Chemical species | PARTRAC[2] | KURBUC[4] | Geant4-DNA[5] | TOPAS-nBio[6] | NASIC |
|----|---|---|---|---|---|---|
| 1 | $e_{aq}^-$ | 4.9 | 4.5 | 4.9 | 4.9 | 4.9 |
| 2 | ·OH | 2.8 | 2.8 | 2.8 | 2.2 | 2.8 |
| 3 | ·H | 7.0 | 7.0 | 7.0 | 7.0 | 7.0 |
| 4 | $H_3O^+$ | 9.0 | 9.0 | 9.0 | 9.46 | 9.0 |
| 5 | $H_2$ | 4.8 | 5.0 | 4.8 | 4.8 | 4.8 |
| 6 | $OH^-$ | 5.0 | 5.0 | 5.0 | 5.3 | 5.0 |
| 7 | $H_2O_2$ | 2.3 | 2.2 | 2.3 | 2.3 | 2.3 |
| 8 | $O_2$ | — | 2.1 | — | 2.4 | — |
| 9 | $O_2^-$ | — | 2.1 | — | 1.75 | — |
| 10 | $HO_2$ | — | 2.0 | — | 2.3 | — |
| 11 | $HO_2^-$ | — | 2.0 | — | 1.4 | — |

Table S2. List of chemical reactions and corresponding observed reaction rate constants ($10^{10}$ M$^{-1}$S$^{-1}$) adopted in different MCTS codes

| ID | Reaction | Products | PARTRAC[2] | KURBUC[4] | Geant4-DNA[5] | TOPAS-nBio[6] | NASIC |
|----|---|---|---|---|---|---|---|
| 1 | ·OH + ·OH | $H_2O_2$ | 0.44 | 0.6 | 0.44 | 0.475 | 0.44 |
| 2 | ·OH + $e_{aq}^-$ | $OH^-$ | 2.95 | 2.5 | 2.95 | 2.953 | 2.95 |
| 3 | ·OH + ·H | $H_2O$ | 1.44 | 2 | 1.44 | 1.438 | 1.44 |
| 4 | ·OH + $H_2$ | ·H | | 0.0045 | | | |
| 5 | ·OH + H$_2$O$_2$ | HO$_2$ | | 0.0023 | | | |
| 6 | ·OH + HO$_2$ | O$_2$ | | 1 | | | |
| 7 | ·OH + O$_2^-$ | O$_2$ + OH$^-$ | | 0.9 | | | |
| 8 | ·OH + HO$_2^-$ | HO$_2$ + OH$^-$ | | 0.5 | | | |
| 9 | e$_{aq}^-$ + e$_{aq}^-$ | H$_2$ + 2OH$^-$ | 0.5 | 0.55 | 0.5 | 0.647 | 0.5 |
| 10 | e$_{aq}^-$ + ·H | H$_2$ + OH$^-$ | 2.65 | 2.5 | 2.65 | 2.652 | 2.65 |
| 11 | e$_{aq}^-$ + H$_3$O$^+$ | ·H | 2.11 | 1.7 | 2.11 | 2.109 | 2.11 |
| 12 | e$_{aq}^-$ + H$_2$O$_2$ | OH + OH$^-$ | 1.41 | 1.3 | 1.41 | 1.405 | 1.41 |
| 13 | e$_{aq}^-$ + HO$_2$ | HO$_2^-$ | | 2 | | | |
| 14 | e$_{aq}^-$ + O$_2$ | O$_2^-$ | | 1.9 | | | |
| 15 | e$_{aq}^-$ + O$_2^-$ | OH$^-$ + HO$_2^-$ | | 1.3 | | | |
| 16 | ·H + ·H | H$_2$ | 1.2 | 1 | 1.2 | 0.503 | 1.2 |
| 17 | ·H + H$_2$O$_2$ | ·OH | | 0.01 | | | |
| 18 | ·H + HO$_2$ | H$_2$O$_2$ | | 2 | | | |
| 19 | ·H + O$_2$ | HO$_2$ | | 2 | | | |
| 20 | ·H + OH− | e$_{aq}^-$ | | 0.002 | | | |
| 21 | ·H + O$_2$- | HO$_2^-$ | | 2 | | | |
| 22 | H$_3$O$^+$ + OH$^-$ | H$_2$O | 14.3 | 10 | 14.3 | 11.031 | 14.27 |
| 23 | H$_3$O$^+$ + O$_2^-$ | HO$_2$ | | 3 | | | |
| 24 | H$_3$O$^+$ + HO$_2^-$ | H$_2$O$_2$ | | 2 | | | |
| 25 | HO$_2$ + HO$_2$ | H$_2$O$_2$ + O$_2$ | | 0.000076 | | | |
| 26 | HO$_2$ + O$_2^-$ | O$_2$ + HO$_2^-$ | | 0.0085 | | | |

The step-by-step simulation algorithm was successfully applied in simulations in oxygen free water[2,3,6,7], however, when extending this algorithm for the oxygenated water, the computation cost will dramatically increase because a huge number of dissolved oxygen molecules need to be considered in simulations. Therefore, an approximated algorithm was adopted in this work to support simulations in oxygenated water.

The approximated algorithm was proposed by Green et al.[8] and Pimblott et al.[9], and adopted in several studies[10-12] for chemical stage simulations in oxygenated water. With such an algorithm, the dissolved oxygen molecules in water can be treated as a continuum and the scavenger of e$_{aq}^-$ and ·H, the time-dependent probability $P(t)$, which is the probability that e$_{aq}^-$ or ·H be scavenged by the continuum in a time duration $t$, can be calculated with:

$$P(t) = 1 - \exp\left(-4\pi D_{sum} R_C C_s \left[t + 2R_C \sqrt{\frac{t}{\pi D_{sum}}}\right]\right) \quad (S3)$$

Where, $D_{sum}$ is the sum of diffusion constants of oxygen and the radical (e$_{aq}^-$ or ·H), $R_C$ is the reaction radius between oxygen and the radical, $C_s$ is the oxygen concentration (mol/L). Therefore, the probability that e$_{aq}^-$ or ·H is not scavenged by the continuum in a time duration $t$ is:

$$Q(t) = 1 - P(t) \quad (S4)$$

Assuming $W(t, \Delta t)$ is the conditional probability that the radical (e$_{aq}^-$ or ·H) is not scavenged by oxygen in the time duration of [0, $t$] and [$t$, $t+\Delta t$], then we have:

$$Q(0, t + \Delta t) = W(t, \Delta t) Q(t) \quad (S5)$$

Thus, the probability that the radical (e$_{aq}^-$ or ·H) is scavenged by oxygen in a time step of $\Delta t$ can be calculated with:

$$P(t, \Delta t) = 1 - W(t, \Delta t) = \frac{P(t + \Delta t) - P(t)}{1 - P(t)} \tag{S6}$$

Therefore, to support chemical stage simulations in oxygenated water, the list of chemical radicals and chemical reactions as well as the diffusion coefficients and the observed reaction rate constants were updated according to Nikjoo et al.[4], and oxygen molecules were not directly introduced in the simulation to avoid a dramatically long computation time, the step-by-step algorithm in combination with the approximated algorithm was used for the simulation:

1. In each time step of $\Delta t$, radicals are diffused and the distances between radicals were scanned to determine if reactions between each pair of reactants happened.
2. If $e_{aq}^-$ or ·H were not reacted in $\Delta t$, then $P(t, \Delta t)$ is calculated and a uniformly distributed random variable $\xi$ is sampled.
3. If $\xi < P(t, \Delta t)$, then it was assumed that radical ($e_{aq}^-$ or ·H) is scavenged by oxygen in $\Delta t$, and the reactant radical is replaced by the product ($O_2^-$ or $HO_2$)

## S2. Validations of NASIC for the simulation in oxygenated water

NASIC was used for the simulation of time-dependent G value, which was defined as the chemical radical yield per 100 eV energy deposition[6], in oxygen free and oxygenated water. Figure S1 and Figure S2 show the comparison of G value simulated with NASIC and reported by Colliaux et al.[11] in oxygen free ($pO_2=0$ mmHg) and oxygenated water ($pO_2=160$ mmHg) after the irradiation of 65 MeV protons and 10 MeV/u carbon ions, respectively. As seen, simulation results obtained with NASIC show good agreements with Colliaux et al.[11].

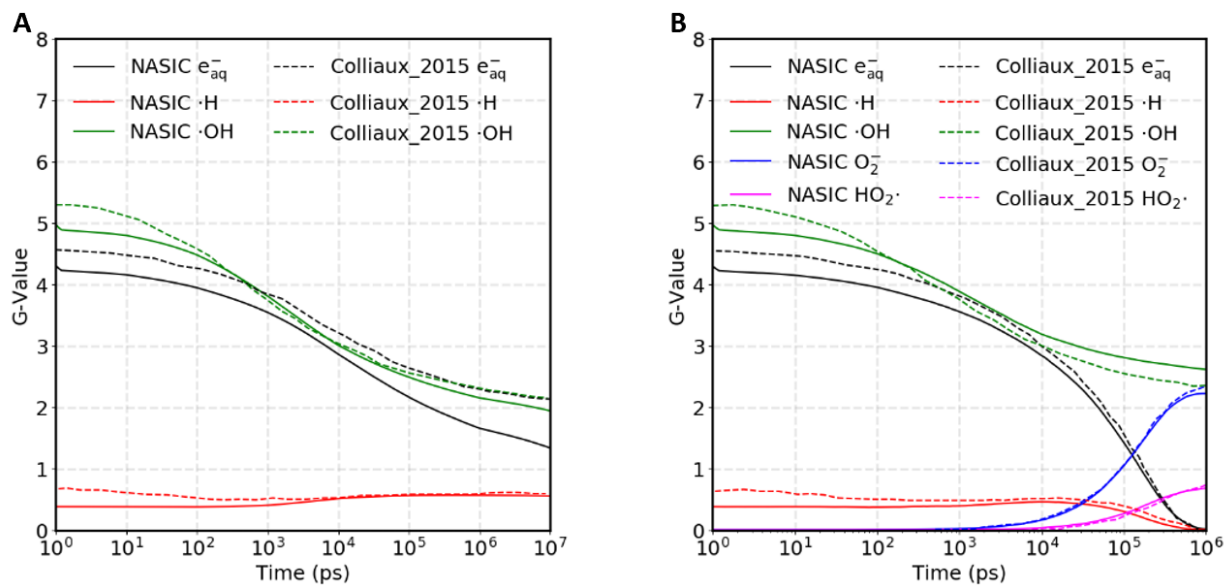

Figure S1. Time-dependent G value simulated with NASIC (solid lines) and reported by Colliaux et al.[11] (dashed lines) in (A) oxygen free water and (B) oxygenated water after the irradiation of 65 MeV protons.

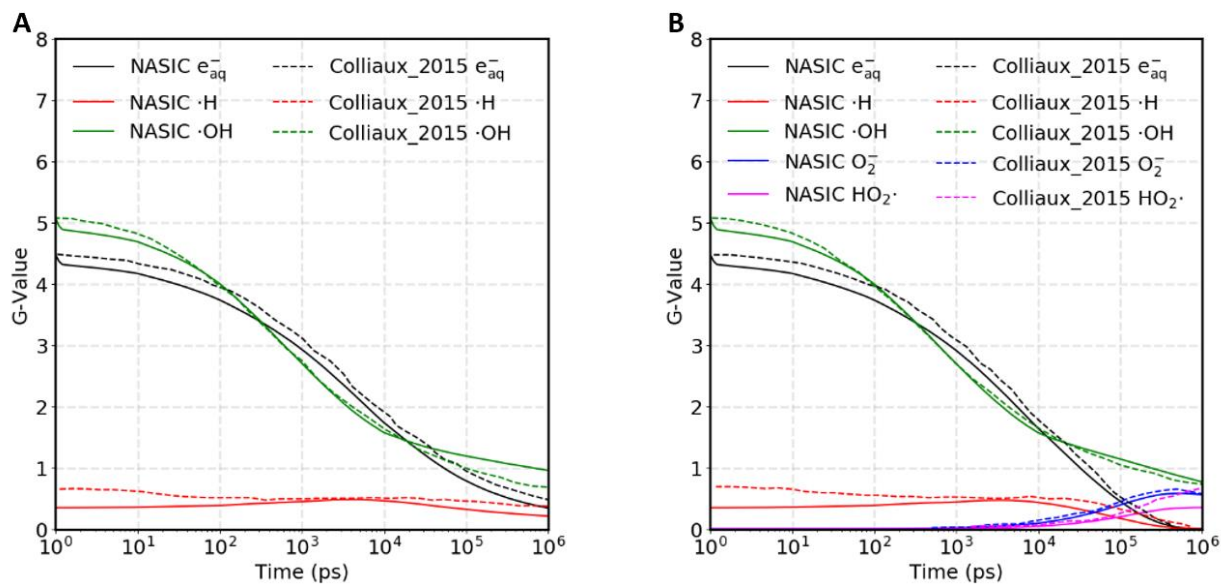

Figure S2. Time-dependent G value simulated with NASIC (solid lines) and reported by Colliaux *et al.*[11] (dashed lines) in (A) oxygen free water and (B) oxygenated water after the irradiation of 10 MeV/u carbon ions.

NASIC was also used to simulate G values in oxygenated water (pre-equilibrated with air, $pO_2$=160 mmHg) 10 μs after the irradiation of $^{60}$Co γ-ray and compared with simulated[11] and experimental[13]. Table S3 shows the results obtained with NASIC are generally in line with experimental results.

Table S3. G values in oxygenated water ($pO_2$=160 mmHg) 10 μs after the irradiation of $^{60}$Co γ-ray

|  | ·OH | $HO_2+O_2^-$ | $H_2O_2$ |
| --- | --- | --- | --- |
| Experimental results by Delattre *et al.*[13] | 2.7 | 3.3 | 0.67 |
| Simulation results by Colliaux *et al.*[11] | 2.8 | 3.4 | 0.61 |
| Simulation results obtained with NASIC | 2.6 | 3.0 | 0.76 |
| Relative difference between NASIC and exp.[13] (1-NASIC/EXP.) | 3.7% | 9.0% | -13.3% |

## S3. The impact of lifetime for $e_{aq}^-$ and ·H radicals on the amount of oxygen depleted in FLASH irradiation

To determine the amount of oxygen depleted in FLASH irradiation ($L_{ROD}$, in units of mmHg/Gy), different radical lifetimes were adopted according to Roots *et al.*[14]. For instance, when the lifetime of ·H and $e_{aq}^-$ was set to 0.19 μs and 40 μs, respectively, all ·H radicals were killed and terminated at 0.19 μs but the other chemical reactions of $e_{aq}^-$ continued to be tracked until the end of its lifetime (40 μs), and the oxygen concentration was recorded for the entire duration (please see Figure S3).

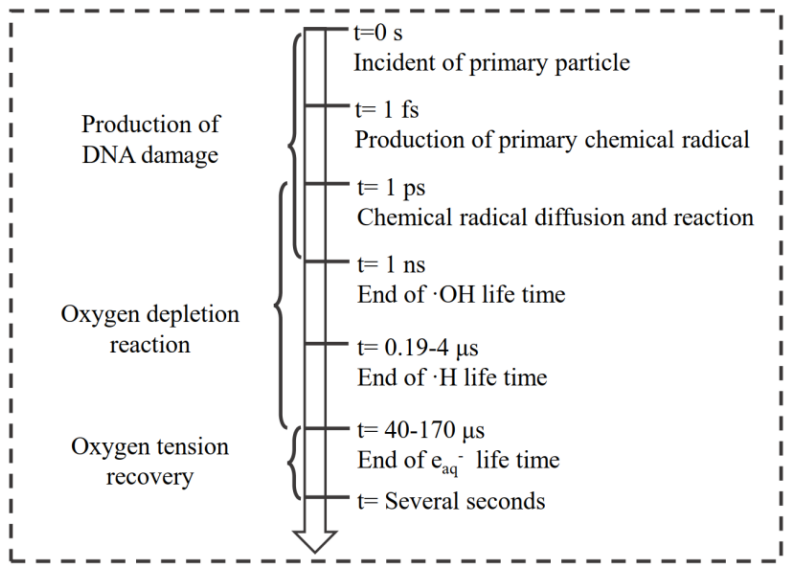

Figure S3. The schematic diagram showing the time scale of radiation-induced reactions. The lifetime values of the radicals in this figure refer to the lifetime in the cellular environment. For MCTS simulations of the cellular responses under CONV irradiation, the simulation is usually terminated at the end of ·OH lifetime. For simulations under FLASH irradiation, the simulation time should be extended to the end of $e_{aq}^-$ lifetime to track the change of oxygen tension due to ROD.

Figure S4 shows the dynamic change of $pO_2$ in cells irradiated by 4.5 MeV electron FLASH pulse with a total dose of 10 Gy. Different radical lifetimes were adopted in the simulations to investigate their impact on the $L_{ROD}$. As seen, $pO_2$ started to decrease from 1 ns, and $pO_2$ decreased faster in cells with a higher initial $pO_2$. The total change in $pO_2$ was greater in cells with a higher initial $pO_2$. Figure S4A shows that a greater change in $pO_2$ was observed when the lifetime of $e_{aq}^-$ was fixed at 40 μs, with ·H having a longer lifetime. However, the impact of the lifetime of ·H on $L_{ROD}$ was limited. This is because the initial ·H yield was approximately 20% of the $e_{aq}^-$ yield and most of the oxygen was depleted due to its reaction with $e_{aq}^-$. Therefore, the changes in the lifetime of ·H did not significantly affect $L_{ROD}$. Figure S4B shows that when the lifetime of ·H was fixed at 1 μs, the change in the lifetime of $e_{aq}^-$ did not affect $L_{ROD}$. This was because even for cells at a low $pO_2$ (15 mmHg), $pO_2$ had been reduced to the lowest level at 10 μs; that is, all $e_{aq}^-$ were scavenged within 10 μs and no more $e_{aq}^-$ was available for chemical reactions after 10 μs. Therefore, increasing the lifetime of $e_{aq}^-$ from 40 μs to 170 μs did not affect the amount of oxygen depleted during FLASH irradiation.

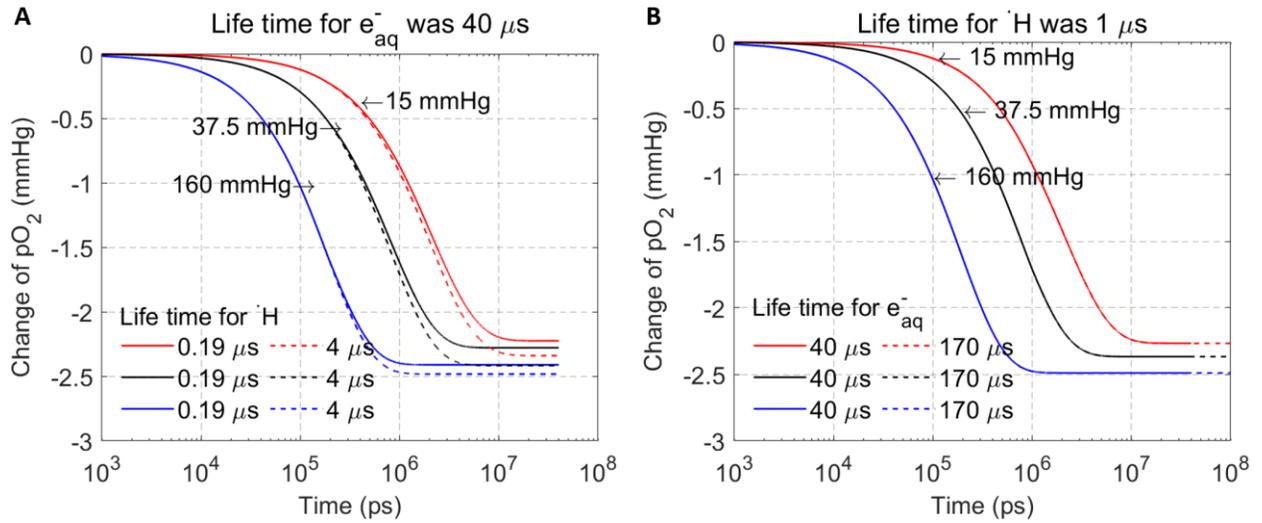

Figure S4. Change in pO$_2$ in cells with a different initial pO$_2$ after being exposed to 4.5 MeV single FLASH pulse with a total dose of 10 Gy. (A) The lifetime of ·H was set to 0.19 or 4 μs, and the lifetime of e$_{aq}^-$ was set to 40 μs. (B) The life time for ·H was set to 1 μs, and the lifetime of e$_{aq}^-$ was set to 40 or 170 μs.

## S4. The DICOLDD (different cell oxygen level DNA damage) model for DNA damage yield calculation

The DICOLDD (different cell oxygen level DNA damage) model[15] developed in our previous study was adopted in this work for the calculation of DNA damage yields in cells at different oxygen tensions. The DICOLDD model was developed based on OFH, and it considers that ionization radiation induces direct DNA damage via physical reactions between DNA and particles, and the indirect DNA damage is converted from DNA radicals produced by chemical reactions between DNA and hydroxyl radicals (·OH). The indirect DNA damage yield changes with oxygen tension in the cell, and the total DNA damage yield, $N_{tot}$, at different oxygen tensions can be calculated as:

$$N_{tot} = N_{dir} + N_{ind} \tag{S7}$$

$$N_{ind} = \frac{0.36 + 25.81 \times C}{0.36 + 25.81 \times C + 20.29\left(1 - \frac{1}{1 + e^{-2.37x + 12.97}}\right)} n_0 \tag{S8}$$

$$x = \ln((Z_{eff}/\beta)^2) \tag{S9}$$

$$Z_{eff} = Z\left[1 - \exp\left(-125 \cdot \beta \cdot Z^{-\frac{2}{3}}\right)\right] \tag{S10}$$

where $N_{dir}$ and $N_{ind}$ are the direct and indirect DNA damage yields, respectively, and $n_0$ is the DNA radical yield. $N_{dir}$ and $n_0$ can be obtained using MC simulations. $C$ is the oxygen concentration in the unit of percentage, $(Z_{eff}/\beta)^2$ is a commonly used parameter that describes the radiation quality of the source[16], $Z_{eff}$ is the effective charge of the source particle, and $\beta = v/c$ is the ratio between the source particle's velocity, $v$, and the light velocity, $c$.

For DNA damage calculations under CONV exposure, it was assumed that the oxygen tension in tissues is at a dynamic equilibrium; that is, pO$_2$ remains unchanged during CONV irradiation, and $C$ in Eq. S8 is the initial oxygen concentration, $C_{ini}$, of cells. For DNA damage calculations under FLASH irradiation, the amount of oxygen depleted ($L_{ROD} \times$ Dose) during irradiation should be simulated, and the value of $C$ in Eq. S8 should be the oxygen tension after irradiation ($C_{ini} - L_{ROD} \times$ Dose).

## S5. The mechanistic cellular survival model for CHO cell survival fraction calculation

In this work, cell SFs were calculated using a mechanistic cellular survival model developed by Wang *et al.*[17]. This model assumes that the double strand break (DSB) is the initial lesion of DNA and that the non-homologous end-joining (NHEJ) pathway is the main DSB repair pathway. It describes the relationship between radiation-induced DSBs in the nucleus and the probability of cell survival. The survival model can be written as Eq. S11 - Eq. S12, and it includes six model parameters ($\mu_x$, $\mu_y$, $\zeta$, $\xi$, $\eta_{\lambda_p \to 1}$, and $\eta_{\lambda_p \to \infty}$), describing the biological characteristics of the irradiated cells. A brief description of the model parameters is given in Table S4.

$$S = exp\left(-\mu_y N \times \left(\frac{1-e^{-\zeta\lambda_p}}{\zeta\lambda_p}\right) \times \left(1 - \mu_x \left(\frac{1-e^{-\xi\lambda_p}}{\xi\lambda_p}\right)\left(\frac{1-e^{-\eta(\lambda_p)n_p}}{\eta(\lambda_p)n_p}\right)\right)\right) \quad (S11)$$

$$\eta(\lambda_p) = \eta_{\lambda_p \to \infty} - \frac{\eta_{\lambda_p \to \infty} - \eta_{\lambda_p \to 1}}{\lambda_p} \quad (S12)$$

In addition to the six model parameters, the model also has two input parameters, $n_p$ and $\lambda_p$, which describe the characteristics of DSB damage. $n_p$ refers to the average number of primary particles that cause DSB and can be calculated using Eq. S13. $\lambda_p$ refers to the average number of DSBs yielded by each primary particle that cause DSB and can be calculated using Eq. S14.

$$n_p = n_{pri}(1 - e^{-\lambda}) \quad (S13)$$

$$\lambda_p = \frac{\lambda}{1 - e^{-\lambda}} \quad (S14)$$

where $n_{pri}$ is the number of primary particles passing through the nucleus in simulations, and $\lambda$ is the DSB yield per cell per primary particle calculated by combining NASIC with the DICOLDD model.

To calculate the SFs of a certain cell line using the cell survival model, six model parameter values should be fitted with experimental data and then used for calculation. Considering the experimental data availability, the Chinese hamster ovary (CHO) cell line was used in this work, and parameter values were fitted using the method described by Wang *et al.*[17] and the data reported by Weyrather *et al.*[18]. The parameter values for CHO cells are given in Table S4.

Table S4. Meaning of model parameters in the mechanistic cellular survival model [17] and the parameter values for CHO cells.

| Model parameters | Parameter meaning | Parameter value for CHO cells |
| --- | --- | --- |
| $\mu_x$ | The average probability of each DSB end correctly joining with the other end from the same DSB via the NHEJ pathway | 0.9888 |
| $\mu_y$ | The cellular sensitivity to DSB mis-repair | 0.0861 |
| $\zeta$ | The parameter that describes over kill effect | 0.1152 |
| $\xi$ | The parameter that describes the effect of clustered DNA damage | 0.0489 |
| $\eta_{\lambda_p \to 1}$ | The parameter that describes the interaction of DSBs induced by different primary particles when $\lambda_p = 1$ | $3.44 \times 10^{-4}$ |
| $\eta_{\lambda_p \to \infty}$ | The parameter that describes the interaction of DSBs induced by different primary particles when $\lambda_p = \infty$ | 0.0021 |

## S6. Cellular responses for normal cells (pO$_2$=15 mmHg) after FLASH irradiation

The impact of FLASH pulse fractionation and pulse interval on the cellular response was investigated, CHO cells (initial pO$_2$ =15 mmHg) were exposed to 4.5 MeV electron FLASH radiation with a total dose of 10 Gy, Table S5 shows a summary of cellular responses with 10 Gy delivered in 1, 2, 5 and 10 fractions, and Table S6 shows a summary of cellular responses with 10 Gy delivered in 10 fractions and each pulse were separated by 10-1000 ms.

Table S5. The average pO$_2$, total DNA damage yield, and cell survival fractions of CHO cells (initial pO$_2$ = 15 mmHg) were exposed to 4.5 MeV electron FLASH irradiation with a total dose of 10 Gy delivered in 1, 2, 5, and 10 pulses; or 10 Gy delivered in 10 pulses which were separated by different intervals.

| Number of pulses delivering 10 Gy | Average pO$_2$ after irradiation (mmHg) | Total DSB yield (Gbp$^{-1}$10 Gy$^{-1}$) | SFs |
|---|---|---|---|
| 1 | 12.74 | 62.28 | 0.072 |
| 2 | 13.81 | 63.13 | 0.068 |
| 5 | 14.20 | 63.42 | 0.066 |
| 10 | 14.16 | 63.39 | 0.067 |
| Maximum relative difference* | 10% | 1.8% | 8.3% |
| Pulse interval (ms) | Average pO$_2$ after irradiation (mmHg) | Total DSB yield (Gbp$^{-1}$10 Gy$^{-1}$) | SFs |
| 10 | 13.77 | 63.09 | 0.0680 |
| 25 | 13.86 | 63.15 | 0.0677 |
| 50 | 13.98 | 63.25 | 0.0672 |
| 200 | 14.39 | 63.56 | 0.0658 |
| 1000 | 14.77 | 63.82 | 0.0646 |
| Maximum relative difference* | 6.8% | 1.1% | 5.0% |

* The maximum relative difference was calculated using: (1 - minimum/maximum.) × 100%